\documentclass[preprint,12pt]{elsarticle}

\usepackage{amsmath}
\usepackage{amssymb}
\usepackage{graphicx}
\graphicspath{{Art/}}
\usepackage[table,xcdraw]{xcolor}
\usepackage{subcaption}
\usepackage{bbold}
\usepackage{float}
\usepackage{csvsimple}
\usepackage{array}
\usepackage{xcolor}
\usepackage{longtable}
\usepackage{makecell}
\usepackage{pgfplots}
\pgfplotsset{compat=1.18}

\usepackage[
]{hyperref}
\usepackage[all]{hypcap}
\hypersetup{
  bookmarksnumbered,
  pdfstartview={FitH},
  citecolor={blue},
  linkcolor={blue},
  urlcolor={black},
  pdfpagemode={UseOutlines}
}

\journal{Journal of Symbolic Computation}

\begin{document}

\begin{frontmatter}



\title{Explainable AI Insights for Symbolic Computation: \\ A case study on selecting the variable ordering for cylindrical algebraic decomposition}

\author[Cincinnati]{Lynn Pickering}
\author[Coventry]{Tereso Del Rio Almajano}
\author[Coventry]{Matthew England}
\author[Cincinnati]{Kelly Cohen}

\affiliation[Cincinnati]{organization={Aerospace Engineering and Engineering Mechanics \\
University of Cincinnati},
            city={Cincinnati},
            postcode={45221}, 
            state={Ohio},
            country={USA}
            }

\affiliation[Coventry]{organization={Research Centre for Computational Science and Mathematical Modelling\\
Coventry University},
            city={Coventry},
            postcode={CV1 5FB}, 
            state={West Midlands},
            country={UK}
            }

\begin{abstract}
In recent years there has been increased use of machine learning (ML) techniques within mathematics, including symbolic computation where it may be applied safely to optimise or select algorithms.  This paper explores whether using explainable AI (XAI) techniques on such ML models can offer new insight for symbolic computation, inspiring new implementations within computer algebra systems that do not directly call upon AI tools.  We present a case study on the use of ML to select the variable ordering for cylindrical algebraic decomposition.  It has already been demonstrated that ML can make the choice well, but here we show how the SHAP tool for explainability can be used to inform new heuristics of a size and complexity similar to those human-designed heuristics currently commonly used in symbolic computation.
\end{abstract}




\end{frontmatter}

\newpage

\section{Introduction}

\subsection{Machine learning and mathematics}
\label{subsec:ml_in_math}

\emph{Machine Learning} (ML) refers to statistical techniques that learn rules from data.  They allow computer systems to improve their performance on a task without any change to their explicit programming.  ML underpins the prominent AI advances of recent years, driven forward by the growth in available data and computing power, new ML architectures, and hardware designed specifically for them.  

There are increasing attempts to use ML in mathematics.  For example, \citet{Lample2020} trained a transformer to integrate expressions and find analytical solutions to differential equations; while in a survey article \citet{He2022} details a variety of attempts to use ML to predict properties from mathematical structures such as groups and graphs.  We note a recent special issue of this Journal of Symbolic Computation dedicated to \emph{Algebraic Geometry and Machine Learning} \citep{HHKMT23} including ML approaches to e.g. find the real discriminant locus \citep{BHMRT23}.  However, it has been observed that mathematical and logical reasoning is an area of particular difficulty for ML, especially the recent natural language tools such as ChatGPT\footnote{See e.g. this blog post by S.~Wolfram: \url{https://writings.stephenwolfram.com/2023/01/wolframalpha-as-the-way-to-bring-computational-knowledge-superpowers-to-chatgpt/}}.

Symbolic computation is not an obvious domain in which to apply ML.  The Computer Algebra Systems (CASs) which implement symbolic computation algorithms are mainly designed to produce exact answers; rarely using numeric computation in preference for symbolic solutions.  \citet{Lample2020} reported that their Transformer\footnote{A particular design of ML model that has found great success in natural language processing tasks: it is used for example by ChatGPT.} was able to integrate far more examples correctly within a time limit than various CASs.  However, this analysis combined two very different cases of failure: when a solution could not be found within a time limit and when the wrong solution is found.  While the CASs may not solve problems quickly, they should \emph{never} produce the wrong answer.  Thus there is little appetite amongst CAS developers to use ML for directly producing output. 

\subsection{Safe use of machine learning in computer algebra systems}
\label{subsec:ml_in_computer_algebra}

This does not mean that ML has nothing to offer Symbolic Computation.  CAS algorithms often come with choices that have no effect on the mathematical correctness of the end result but can have a big impact on the resources required.  Further, they can affect how the end result is presented (i.e. two very different, but mathematically equivalent, expressions).  
Consider for example the algorithm of \citet{Buchberger2006} to produce a Gr\"{o}bner Basis for an ideal.  It does not specify the order in which $S$-pairs are computed, the order in which the corresponding $S$-polynomial is reduced by the generating set, the monomial ordering to be used, nor the underlying variable ordering.  Any decision for these choices allows the production of a correct Gr\"{o}bner Basis but each decision affects the size of the basis produced, the polynomials within, and the time taken to compute them.  

Sometimes there are well-documented strategies on how CASs make such choices (e.g. sugar selection for $S$-pairs \citep{GMNRT91}) but in many cases, they rely on human-made heuristics or ``\emph{magic constants}'' \citep{Carette2004} which are often not scientifically validated or even documented.  Such choices are good candidates for ML:  the underlying relationships are complex but not the main object of study.  The earliest examples of ML for CAS optimisation known to the authors are: in 2014 \citet{Huang2014} used a support vector machine to choose the variable ordering for cylindrical algebraic decomposition; in 2015 \citet{KUV15} used a Monte-Carlo tree search to find the representation of polynomials that are most efficient to evaluate; and in 2016 \citet{SYK16} used ML classifiers to pick from various algorithms that compute the resultant, \citet{HEDP16} used ML to decide whether to precondition input for cylindrical algebraic decomposition, and \citet{Kobayashi2016} used ML to decide the order of sub-formulae solving for real quantifier elimination.

In all of these examples, the ML solution outperformed the previous human-made heuristics for the choices, at least on the dataset used.  The algorithms of symbolic computation are often exponential in their worst-case complexity but the average case complexity is less widely studied.  It is possible that these ML optimisations may offer a route to avoid the worst case, at least for many applied examples.  Thus there is great value in their continued study and exposition.  However, the motivation of the present paper is whether ML can offer anything \emph{more} than these efficiency gains.  A recent article in Nature \citep{DVBBZTTBBJLWHK21} suggested that ML can help pure mathematicians with the development of new theorems.  Can ML offer such insight to the developers of symbolic computation algorithms?

\subsection{Explainable artificial intelligence}
\label{subsec:explainable_ai}

Explainable AI (XAI) refers to ML techniques which offer an explanation as to how an AI decision was made.  It has become a very active area of research, see for example DARPA's XAI program \citep{gunning_darpas_2019}.   
The role of XAI is usually to allow for more effective use of the AI tools, and the building of user trust in the tools.  But in our case, we hypothesise that XAI could reveal workings of the ML model that are useful to inform further development of the underlying symbolic computation.

We were motivated by the work of \citet{Peifer2020} developing an ML solution to the problem of selecting $S$-pairs in Buchberger's algorithm. They trained an ``\emph{agent}'' (ML model) to make the decision using reinforcement learning\footnote{With reinforcement learning, instead of having a labelled dataset to train against, the agent repeatedly makes decisions which receive a score that they seek to optimise.} based on the number of polynomial additions required to process the choice.  This agent outperformed existing strategies such as \citep{GMNRT91}.  We draw attention to \citep[Section 5.1]{Peifer2020} which tried to explain the agents' strategy (albeit without employing XAI tools).  It identified some simple components such as a preference for pairs whose $S$-polynomials are monomials and a preference for pairs whose $S$-polynomials are low degree.  Firstly, these components suggested decisions made on the basis of the $S$-polynomials rather than the $S$-pairs\footnote{even though it was properties of the pairs fed to the network.}, which is what all the human-made heuristics consider.  Hence this opens a whole new category of heuristics to consider, perhaps leading to a greater understanding of the problem.  Secondly, experiments using these strategies alone showed they outperformed the existing heuristics (but not the full ML agent).  Thus the analysis suggests new ``\emph{human-level}'' heuristics, i.e. heuristics of a similar size to those that are human-made and that, once discovered, can be implemented without any AI tools.

The present paper seeks to identify if XAI tools can offer an automated way to make such analyses, to gain new understanding of these heuristic choices in symbolic computation algorithms.

\subsection{Contributions and plan of the paper}

In this paper, we return to the problem of selecting the variable ordering for cylindrical algebraic decomposition, the first CAS optimisation by ML that was considered in \citep{Huang2014}.  We introduce the necessary background information on this problem in Section \ref{sec:CAD}.  We then apply the popular SHAP methodology for XAI to an existing ML pipeline developed for the problem: we introduce SHAP in Section \ref{sec:XAI} and report on the results of its application on this problem in Section \ref{sec:XAI_CAD}.  In Section \ref{sec:explainability_analysis} we process the SHAP data to suggest some XAI-recommended features for the problem, and in Section \ref{sec:new_heuristics} we report on the performance of human-level heuristics based upon these.  The code and data underpinning these results are openly available from Zenodo here:  \url{https://doi.org/10.5281/zenodo.8229298} 

We find that one of the features recommended by XAI is the same as that used in the current state-of-the-art human-developed heuristic for the problem \citep{DelRio2022}, and further, that a human-level combination of XAI selected features allows us to outperform that prior work.  Thus our contributions are (a) a new state-of-the-art human-level heuristic, at least in three variable problems, and perhaps more importantly; (b) evidence for a new methodology to employ in CAS development that uses AI in algorithm design but not implementation.

\section{Cylindrical Algebraic Decomposition}
\label{sec:CAD}

\subsection{Cylindrical algebraic decomposition}
\label{subsec:cad}

Cylindrical Algebraic Decomposition (CAD) is an important symbolic computation algorithm, proposed by \citet{Collins1975} in 1975.  Given a set of multivariate polynomials and a variable ordering, CAD will decompose the corresponding real space into connected regions, called cells, such that within each the sign of each polynomial is invariant (i.e. positive, negative or zero). Each cell is semi-algebraic, meaning it may be defined by a set of polynomial constraints; and the cells are arranged in cylinders with respect to the variable ordering, meaning the projections of any two cells onto a lower dimensional space with respect to the ordering are either equal or disjoint.  

CAD has found applications in various fields, ranging from robotics \citep{MMCRW12} through economics \citep{MDE18} to biology \citep{Rost2021}.  It is most commonly used for, and was proposed by \citet{Collins1975} as,  a sub-method to perform real Quantifier Elimination (QE): i.e. to transform a formula in first-order logic whose atoms are Boolean constraints to an equivalent one without any quantifiers.  The decomposition into cells allows for studying the problem at a finite number of points, while the cylindrical structure allows for easy projection and negation of the cells to construct the quantifier-free solution.  

However, CAD has a worst-case theoretical complexity that grows doubly exponential in the number of variables \citep{Davenport1988}; something regularly encountered in practice and thus vastly reducing the scope of use cases for the algorithm. There have thus been many lines of research to improve CAD, such as improved projection operators \citep{McCallum2019}, partial CAD \citep{CH91}, equational constraints \citep{EBD20}, symbolic-numeric lifting \citep{Strzebonski2006}, alternative computation schemes via triangular decomposition \citep{CMXY09} and incremental processing of polynomials \citep{KA20}, and entirely new repackaging of the CAD theory \citep{Brown2015}, \citep{Jovanovic2012}.

\subsection{CAD variable ordering}
\label{subsec:variable_ordering}

CAD, including all the optimised and extended forms listed above, requires there to be a declared ordering on the variables.  This controls both the order of operations in the algorithms and  the form of the output (being used in the definition of the cylindrical structure).  Depending on the application, there may be a free or constrained choice for the ordering.  For example, when using CAD for real QE the variables must be ordered as they are quantified: but variables in blocks of the same quantifier (and free variables) can be swapped with each other.  

\begin{figure}[t]
	\centering
	\includegraphics[width=0.48\textwidth]{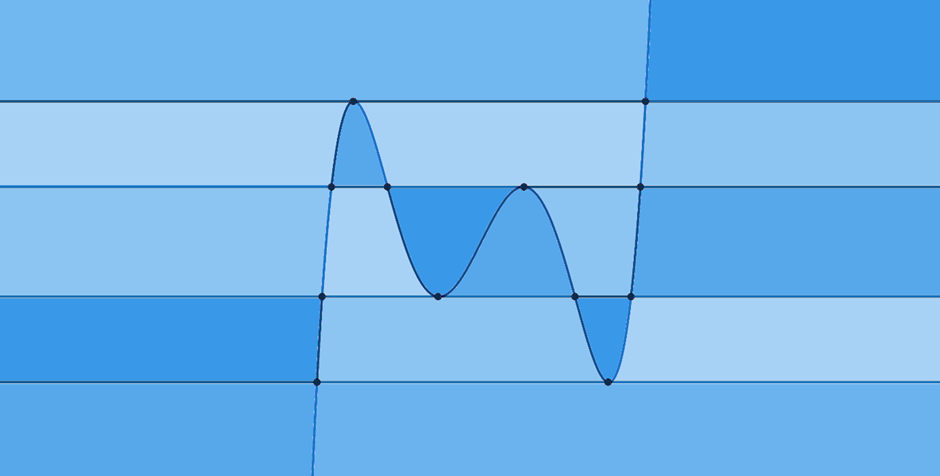}
    \hspace{0.1pt}
	\includegraphics[width=0.48\textwidth]{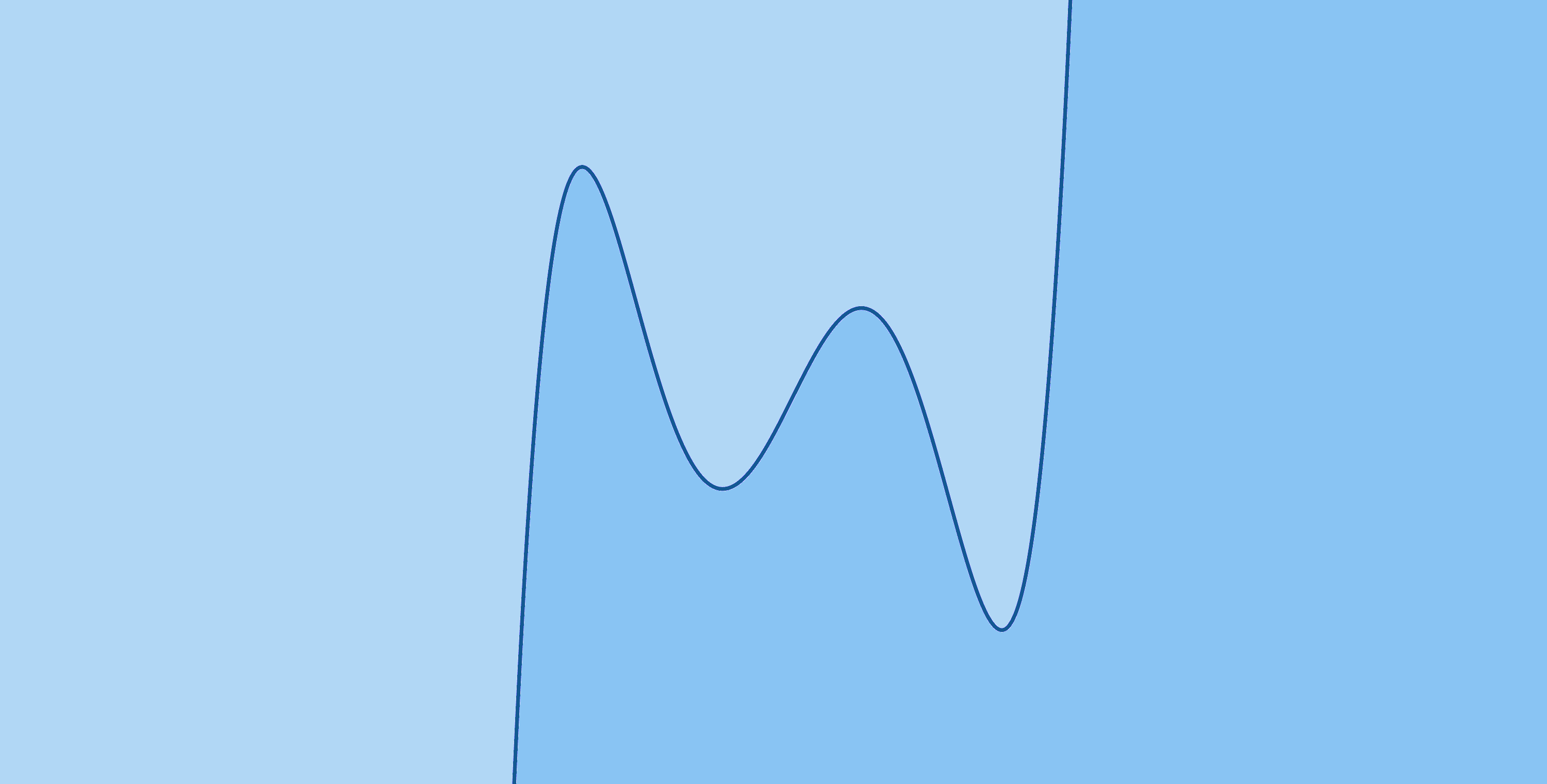}
	\caption{CADs sign-invariant for the set of polynomials $\{x^5+5 x^4+5 x^3-5 x^2-6x-2y\}.$
            Using ordering $x\succ y$, we obtain a CAD with 57 cells (18 shaded areas, 27 curve segments and 12 points).
            Using the ordering $y\succ x$ generates only 3 cells (2 areas and one curve).  
            \label{fig: importanceordering}
            }
\end{figure}

This choice of variable ordering has been shown to have a huge impact, both practically \citep{Dolzmann2004} and in terms of theoretical complexity: \citet{Brown2007} showed there exists a family of polynomial sets such that the complexity of CAD can be constant or doubly exponential depending on the chosen variable ordering. 
It is possible to appreciate the importance of variable ordering in the simple two-variable example of Figure \ref{fig: importanceordering}, where using one of the orderings generates almost twenty times more cells than using the other.

\subsection{Prior human-designed heuristics for choosing the ordering}
\label{subsec:prior_heuristics}

Once the importance of variable ordering was established, researchers began looking for strategies to choose a good variable ordering. The first attempts consisted of designing human-made heuristics. In 2004 \citet{Brown2004} documented a heuristic based on three simple hand-picked features of the set of polynomials, for the software \textsc{QEPCAD}. We call this heuristic \texttt{Brown}.

There have been other heuristics produced which can offer greater accuracy but at greater expense, performing increasing numbers of steps in the CAD algorithm:  \citet{Dolzmann2004} concluded it is best to perform the projection stage of CAD and compare sums of the total degree of the polynomials produced (\texttt{sotd}); \citet{BDEW13} considered the initial decomposition of the real line; and \citet{Wilson2015} the open cells in the decomposition.  
It is perhaps surprising that the original heuristic of Brown is able to perform competitively or even beat these others despite having access to less information \citep{Huang2019}.  This suggests the necessary information to identify a good ordering may be available from the input alone.

Most recently, \citet{DelRio2022} designed a new simple heuristic \texttt{gmods} based on properties of the input polynomials, selected by studying which features of the input polynomials have the greatest impact on the complexity analysis of CAD.

\subsection{Prior AI-designed heuristics for choosing the ordering}
\label{subsec:prior_ml_models}

In the last decade machine learning models have also been trained to select the variable ordering for CAD.  The first attempt was made by \citet{Huang2014} in 2014 who used a support vector machine for the task.  Later \citet{EF19} experimented with a wider range of models, methods for feature engineering \citep{Florescu2019a} and improved metrics for training \citep{florescu2020improved}, culminating in a machine learning pipeline available to use for the task \citep{florescu2020machine}.  Separately \citet{CZC20} experimented with deep learning for variable selection.

In all the experiments performed in this paper, these ML models outperform existing heuristics.  However, CAS developers may still be reluctant to include a method trained using ML in their software, possibly because they create a dependence on external code, or more likely due to concerns about whether over-fitting on the training data will lead to poor generalisation.   Another issue is the lack of explanation these AI models are able to give about their choices. 

Later, in Section \ref{sec:XAI_CAD}, we will use the infrastructure for a ML selection of the CAD variable ordering presented in \citep{florescu2020machine} in our own XAI analysis. In that paper, the authors present a pipeline to train four ML models on the task.  When specified to three variables polynomials the pipeline utilises the ideas described in \citep{Florescu2019a} to represent such polynomials with 81 features (which evaluate to floating point values).  It is these numerical feature vectors that are presented to the ML models.

In this paper, we propose an approach that is somewhere in between those of this subsection and the previous one.  Rather than handpicking some features as in Section \ref{subsec:prior_heuristics}, we will start with the algorithmically produced features of Section \ref{subsec:prior_ml_models} but then use XAI techniques to extract the most important features from the AI model and use these to design heuristics that could easily be understood by humans and embedded into any CAD implementation without the subsequent use of ML software.

\section{Explaining Machine Learning Models}
\label{sec:XAI}

XAI is often motivated as a means to build trust in the tool among users.  But there are certainly other reasons we may require an explanation of an ML model.  Perhaps the model is making a poor decision and the data scientist wishes to understand why so as to improve it; or perhaps the explanation tells us more about the underlying trends of data leading to new research directions. The latter is the motivation for explainability in this work. Within the XAI field there is a growing emphasis on designing models that are intrinsically explainable\footnote{There are well-established views on which models are intrinsically explainable and an assumed tradeoff against performance with e.g. neural networks offering the best performance but the least explanation.  However, in practice this trade-off is not always so clear, see e.g. \citep{HHWJ22}.}, however, the most common approach in use today is post-hoc explainability whereby a model is analysed after its creation and an explanation is generated.  The review article by \citet{DLH19} offers a good overview of XAI.   

Since we work with the existing ML pipeline \citep{florescu2020machine} already designed for the problem we do not consider editing the ML itself for explainability but instead revert to the more common post-hoc analysis.  That pipeline works with several ML models, and so we desire a model-agnostic method for explainability. The following is the current state-of-the-art process for this.

\subsection{SHAP} 
\label{subsec:shap}

\citet{molnar_interpretable_nodate} gives an overview of many explainability methods along with their advantages and disadvantages. The choice of explainability method will have an impact on what conclusions may be drawn.  
SHAP (SHapley Additive exPlanations), a model agnostic method, \citep{lundberg2017unified} is chosen as our method of post-explainability analysis for identifying the features that have a greater impact on the output of the ML models. The SHAP approach was introduced in 2017 by \citet{lundberg2017unified} as a unification of six existing explanation methods already present in the literature: LIME, DeepLIFT, Layer-Wise Relevance Propagation, Classic Shapley Regression Values, Shapley Sampling Values, and Quantitative Input Influence. Because SHAP unifies these various methods, it can be used for both a global and local explanation of models: i.e. it can offer explanations for performance on an individual problem or an entire dataset. A theorem in \citet{lundberg2017unified} shows that there is only one possible additive feature attribution method (SHAP) that holds local accuracy, missingness (a missing feature must have no impact) and consistency properties. 

The name and theory behind SHAP is based on the classical concept of Shapley values from game theory, which quantifies the contribution of each player to a game \citep{Shapley195317AV}. It makes use of an explainer model, which is an approximation of the original model. The idea is to figure out the contribution of each player to the score of the game by playing the game with all the combinations of players, or the power set of the players. Take an example game with three players: Player 1, Player 2, and Player 3. Each player plays the game by themselves, then each player plays the game with each of the other two, and finally they all play the game together. We calculate the difference between the final scores of each game where Player 1 was not present, and when Player 1 was present under a weighting to get the contribution of Player 1 to the game.  In our application of SHAP, the features of our polynomial sets are the players and the ML models are the games. SHAP removes and adds each feature to the model to see what effect that feature has.  

Since we are working over the power set of all the features in the model, such a calculation becomes very expensive quickly: exponential on the number of features in a model \citep{lundberg2017unified}.  However, the SHAP open source package \citep{shap_git} uses many estimators and powerful shortcuts to make the extensive power set of calculations possible for larger models. For example, Kernel SHAP, used to analyse all the models other than decision tree in this work, uses a special weighted linear regression in computing a feature's importance \citep{lundberg2017unified}. For the decision tree model, the tree explainer is used. This is a fast method to estimate SHAP values for tree models \citep{lundberg2020trees}.

\subsection{SHAP waterfall plots}
\label{subsec:waterfall}

\begin{figure}[t]
     \centering
     \includegraphics[width=0.9\textwidth]{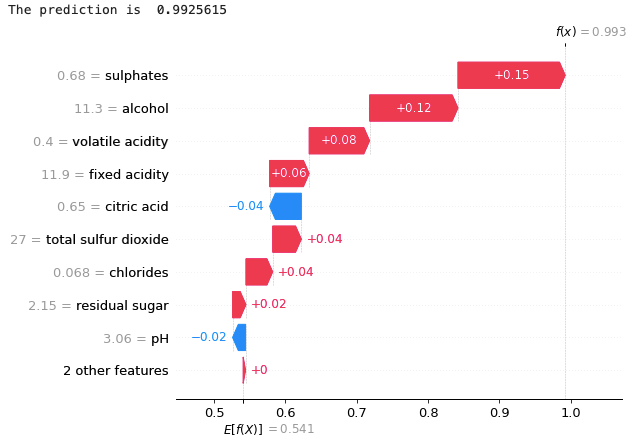}
    \caption[]{SHAP waterfall plot reproduced from an online tutorial (\url{https://medium.com/dataman-in-ai/the-shap-with-more-elegant-charts-bc3e73fa1c0c}) where it appears as the second image in Section 3.2.  The plot explains the prediction by an XGBoost model for whether a particular wine has a rating in the top half of the quality scale, using the ``\emph{Red Wine Quality}'' Kaggle  dataset (\url{https://www.kaggle.com/datasets/uciml/red-wine-quality-cortez-et-al-2009}).    
    \label{example_shap}}
\end{figure}

Figure \ref{example_shap} shows an example of a SHAP waterfall plot: one visualization in SHAP for understanding a model's decision on a single instance. The ML model studied here was trained to decide whether a wine has a quality rating higher than 5 (on a scale of $1-10$).    

The horizontal axis plots the probability of the wine having quality higher than 5 on the scale.  We see that the model predicts this with probability 0.993 for this instance (this particular wine), indicating a strong predicting that it will be ranked in the top half of the quality scale.  

The features that contribute to a prediction that a wine has a quality rating strictly over 5 are the red bars, while the features that contribute to a prediction that a wine has a quality rating of 5 or under are the blue bars. The features from top to bottom are those that have the largest effect on the final model output.  Thus we see that, for this example, \emph{sulphates} (a measure of the presence of this wine additive) is the feature with the most impact on this prediction. The instance from the data set under study had 0.68 value for \emph{sulphates} as indicated in gray on the vertical axis: the value of this feature in the whole dataset ranged from $0.33-2$.

Below the horizontal axis we see $E[f(X)]$ which is the base value for this probability.  It is the average of the output across all the instances, which would be the best guess a model would have for this dataset if there were no features for the instance to consider. Thus the waterfall plot explains the difference between a particular prediction $f(x)$, and this base prediction $E[f(x)]$ for the dataset, and how each feature contributed to this difference.

\subsection{Multi-class SHAP}
\label{subsec:multiclass_shap}

The ML pipeline we experiment with is designed for multi-classification, i.e. classification with more than two output classes. SHAP may also be applied for multi-classification.  It simply creates a binary model for each output, with the binary SHAP model that reaches an output closest to one for an instance giving the prediction. Section \ref{sec: local shap analysis} provides a more detailed example of this in practice.  

\section{SHAP Applied to ML for CAD Variable Ordering}
\label{sec:XAI_CAD}

We work with the feature engineering approach to the problem developed by \citet{Florescu2019a}.  This algorithmically generates features to represent systems of polynomials to ML models.  It was inspired by the earlier hand-picked features based on variable degree and sparsity \citep{Brown2004}, \citep{Huang2014} and generalises this to the widest set of such features that could be generated in a similar way.  When applied to three variable polynomial systems this framework creates 81 distinct features, which the pipeline in \citep{florescu2020machine} uses as the input to train four ML models (implemented in Python from the scikit-learn package \citep{scikit-learn}) which are: Decision Tree (DT), K-Nearest Neighbors (KNN), Support Vector Machine (SVM), and Multi-Layer Perceptron (MLP)\footnote{We refer a reader seeking the details of these models to e.g. the textbook \citet{Bishop2006}.}. 

We fit model hyperparameters using the same cross-validation procedure as in \citep{florescu2020machine}. The DT uses a random splitter with a max depth of 5, and the gini criterion to measure the quality of a split in the tree. The KNN uses the ball tree algorithm to find the 18 nearest neighbors and distance to weight the points in the neighborhood. The SVM has a gamma of 0.08, and the radial basis function kernel. The MLP has logistic activation, hidden layer sizes of 2, and the lbfgs solver\footnote{We refer a reader seeking the details of these implemented hyperparameters to the scikit-learn documentation for each respective model as linked from: \url{https://scikit-learn.org/stable/supervised_learning.html}}. The output that each model predicts is a class (1$-$6) representing one of the six possible variable orderings of three variables. In these papers, the experiments showed the ML models to all outperform the best performing human-made heuristic of the time, that of \citet{Brown2004}.  Although \citep{Brown2004} has since been outperformed by the heuristic \texttt{gmods} obtained by \citet{DelRio2022}. 

The work described in this section uses SHAP to gain insight into how those ML models from \citep{florescu2020machine} achieve this performance.  We seek to understand whether the ML models find the most important features to be those selected by humans, which are then supplemented with additional information; or whether they instead rely on some previously overlooked feature. This section first looks at the decision-making process for a single test instance to gain a local understanding of a particular ML model (the MLP). The process from the local decision to the global decision is then explained, and the top five features across the entire model, for each model are identified and discussed.

\subsection{Methodology}
\label{subsec:Methodology}

The methodology used in applying SHAP to ML algorithms for picking the CAD variable ordering is described next, starting with the data set used to train the algorithms, the specific manner in which SHAP was implemented, and details on the features used as input to the ML algorithms. 

\subsubsection{Dataset} 

We started by recreating the experiments of \citet{florescu2020machine}.  These used a dataset of random problems generated with the same statistical properties as the industrial problems hosted in the QF\_NRA category of the SMT-LIB \citep{SMTLIB}.  We identified that the dataset used in \citet{florescu2020machine} had a majority of problems with class 5 or 6 as the output class.  In this paper, we applied the additional step of balancing the dataset so that the classes are more equal in size. We take advantage of the fact that the nature of a polynomial set does not change with the names of its variables. This allows for the random permutation of variable names $(x_1, x_2, x_3)$, and a corresponding change of the best ordering for the output, to obtain a balanced version of the dataset. As a reference of which output class number corresponds to which order of the variables, we refer to Table \ref{tab: orderings}. 

\begin{table}[ht]
\def\arraystretch{1.5}
\setlength\tabcolsep{0.2cm}
\centering
\caption{The six possible variable orderings for our dataset}
\label{tab: orderings}
\begin{tabular}{|c|c|}
\hline
\multicolumn{1}{|l|}{\textbf{Ordering Name}} & \textbf{Ordering } \\ \hline
Ordering 1  & $x_1 \succ x_2 \succ x_3$    \\ \hline
Ordering 2  & $x_1 \succ x_3 \succ x_2$    \\ \hline
Ordering 3  & $x_2 \succ x_1 \succ x_3$    \\ \hline
Ordering 4  & $x_2 \succ x_3 \succ x_1$    \\ \hline
Ordering 5  & $x_3 \succ x_1 \succ x_2$    \\ \hline
Ordering 6  & $x_3 \succ x_2 \succ x_1$    \\ \hline
\end{tabular}
\end{table}

We note that while this renaming makes no difference to the meaning of the dataset, training an ML model on an unbalanced dataset can lead to the ML model simply choosing the class that occurs the most often, and being rewarded for higher accuracy in this way.  In general, we might expect this to lead to overfitting, where the model may perform far worse on data from a different source.  In our experiments the Decision Tree was found to have particularly suffered from this, making far poorer choices when trained with the balanced dataset, however, the other three models performed similarly well with balanced and unbalanced data and in particular, still outperformed the human-designed heuristics in the unbalanced case.  The effect of such balancing on performance is explored in more detail in \cite{dRE23}.

From the point of view of this paper, balancing is particularly important for explainability.  Our initial SHAP analysis on the models trained on the unbalanced dataset clearly reflected a simplistic preference for the majority classes and gave us little insight into understanding the important features of this application.

\subsubsection{SHAP}

SHAP is run on the four ML models. Table \ref{tab: shap_times} shows the very significant differences in SHAP run-time between the models. This is due to both the nature of how long a single ML model takes to calculate a prediction, and the approximations that the SHAP library is able to make use of. We note that SHAP is a procedure we are running once to gain insight, and so are not greatly concerned with its one-off cost.  
Due to the long-running time for SHAP, 500 training points are used for each SHAP calculation, chosen randomly from the 2947 total training instances used to train the full models described above, and the SHAP calculations are performed on just 100 test points.  
This gives sufficient insight into the model decisions without requiring more than 12 hours of calculation time. We refer back to Section \ref{subsec:shap} for information on how SHAP makes its calculations.

\begin{table}
\def\arraystretch{1.5}
\setlength{\tabcolsep}{10pt}
\centering
\caption{Wall clock runtimes of the SHAP analysis for each model with 500 training points and 100 test points. CPU used is Intel(R) Xeon(R) CPU E5-1620 v3 @ 3.50GHz}
\label{tab: shap_times}
\begin{tabular}{| c | c| }
\hline
\textbf{Model} &  
    \begin{tabular}{@{}c@{}} \textbf{Total Run Time} \\ \textbf{(HH:MM:SS)}\end{tabular} \\
\hline
KNN & 07:02:42\\
\hline
MLP & 00:09:52\\
\hline
DT & 00:00:10\\
\hline
SVM & 11:38:43\\
\hline
\end{tabular}
\end{table}

\subsubsection{Features}
\label{subsubsec:features}

The derivation of the features used in the dataset is explained in \citep{Florescu2019a}. In brief, instead of handpicking a set of features we algorithmically generate multiple features.  This is done by acting upon a list of lists of lists with the innermost list capturing the degrees of variables in a monomial, which are gathered together next in polynomials, and then finally for the whole system.  The actions performed on these are simple (cheap) operations such as taking the maximum, sum or average of the numbers.  By looking at all possible combinations this method generates 81 distinct features for three variable systems that are the input features for the ML models. We define in Table \ref{tab: feature_translation} some of the terminology used in creating these features.  

\begin{table}[H]
\def\arraystretch{1.5}
\setlength\tabcolsep{0.2cm}
\centering
\caption{Translation of Features}
\label{tab: feature_translation}
\begin{tabular}{|c|l|}
\hline
\multicolumn{1}{|l|}{\textbf{Notation}} & \textbf{Text Translation} 
\\ \hline
$S$  & Lists of polynomials.  
\\  & E.g. $S=[x_1,x_1^2-2x_1x_2^2+x_2^2-3]$.         
\\ \hline
$v_i(S)$ \vspace{-5pt} & A list that has for each polynomial in $S$ a list 
\\ & of the degrees of $x_i$ in its monomials. 
\\   & E.g. $v_2(S)=[[0],[0,2,2,0]]$.
\\ \hline
$sv_i(S)$ \vspace{-5pt} & A list that has for each polynomial in $S$ a list  \vspace{-5pt} \\ & of the total degrees in its monomials if $x_i$ appears  \vspace{-5pt} \\ & in the monomial, and $0$ otherwise.
\\   & E.g. $sv_1(S)=[[1],[2,3,0,0]]$.
\\ \hline
$op(L)$                         & If $L$ is a list it performs operation $op$ upon it. \vspace{-5pt}\\ &If $L$ is a list of lists it returns a list containing the  \vspace{-5pt} \\ & the result of the operation for each of the original lists.\\ & E.g. $max(avg(sv_1(S)))=max([1,5/4])=5/4$. 
\\ \hline
$sg(L)$   \vspace{-5pt}                       & This returns the same structure as $L$ where each 
\\ & numerical value is substituted by its sign.  
\\ & E.g. $sum(sg(avg(sg(v_2(S)))))=sum(sg([0,1/2]))$ \vspace{-5pt} 
\\ & $\hspace{155pt}=sum([0,1])=1$.  
\\ \hline
\end{tabular}
\end{table}

For example, $sum( max( v_1(S)))$ is one of the features that multiple models place importance upon.  Its direct translation into plain English is 
``\textit{the sum across all polynomials of the maximum of the degrees of its monomials for the variable $x_1$}'',
i.e. $\sum_{p}\max_{m \in p}degree(x_1,m)$ where $m$ represents monomials and $p$ polynomials.  This simplifies to, ``\textit{the sum across all polynomials of their degrees for the variable $x_1$}'', i.e. $\sum_{p}degree(x_1,p)$.

Similarly, the meaning of $avg(sg(max(v_1(S))))$ is ``\textit{the average across all polynomials of the sign of the maximum of the degrees of its monomials for the variable $x_1$}'', which simplifies to  ``\textit{the average number of polynomials that contain the variable $x_1$}''.

\subsection{SHAP Results}

In this section, we interpret the results from SHAP when applied to our problem using the methodology above.  First, we explain how it can be used to give an explanation for the classification of a single problem instance:  a local SHAP analysis. Then we aggregate across all instances to give an overview of what SHAP tells us about the most relevant features for each model over the dataset: a global SHAP analysis. 

\subsubsection{Local SHAP Analysis}
\label{sec: local shap analysis}
\begin{figure}
 \centering
 \includegraphics[width=0.85\textwidth]{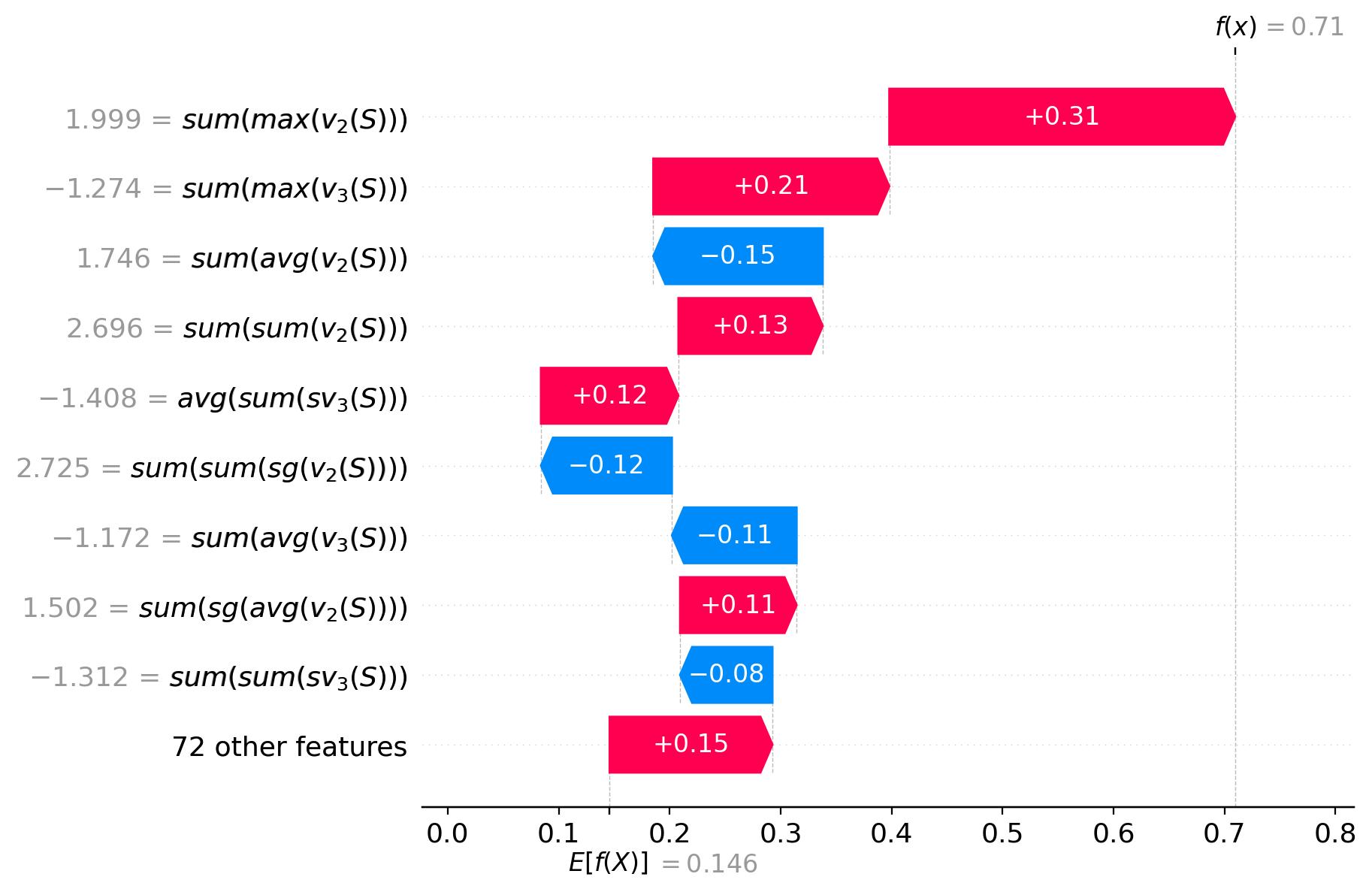}
\caption{An explanation of a decision made by the MLP model on an example CAD problem instance, for the selected output ordering, ordering 5: $x_3 \succ x_1 \succ x_2$}
\label{fig: waterfall_ordering}
\end{figure}

For each ML method, for each of the six binary classification models, and for each instance, SHAP can produce a waterfall plot (see Section \ref{subsec:waterfall}) to show the effect each feature has on the final output. As an example, let us look in detail at the MLP model for a particular test instance.  An explanation for the binary classifier of whether or not the instance belongs to class 5 (selects Ordering 5) is given via the waterfall plot in Figure \ref{fig: waterfall_ordering}.  
The blue arrows in the waterfall indicate that the feature is pushing the prediction towards an output of 0 (that Ordering 5 is the wrong ordering), and the red arrows that the feature pushes towards an output of 1 (that Ordering 5 is the right ordering). 

The binary classifier suggested the instance should be classified as Ordering 5 with a probability of 0.71. This was the highest probability among the six binary classifiers and so Ordering 5 was selected as the classification.  Orderings $1-4$ get a score very close to 0. Ordering 6 receives a score much higher than 0, but not as high as Ordering 5. This is not surprising: Orderings 5 and 6 both select the same variable to project first and it is sensible to assume that this would have the greatest impact on the calculation time of the CAD, since its effects will be magnified in the subsequent projections. 

\subsubsection{Global SHAP Analysis}

To get the global importance of the features for each model over the dataset we may aggregate the absolute values of the contribution of the features across all instances and binary classifiers.  After this, the five most important features for each model found by SHAP are given in Table \ref{tab: top5_all_models}. 

\begin{figure}
 \centering    
     \begin{subfigure}{.59\textwidth}
         \centering
         \includegraphics[width=\textwidth]{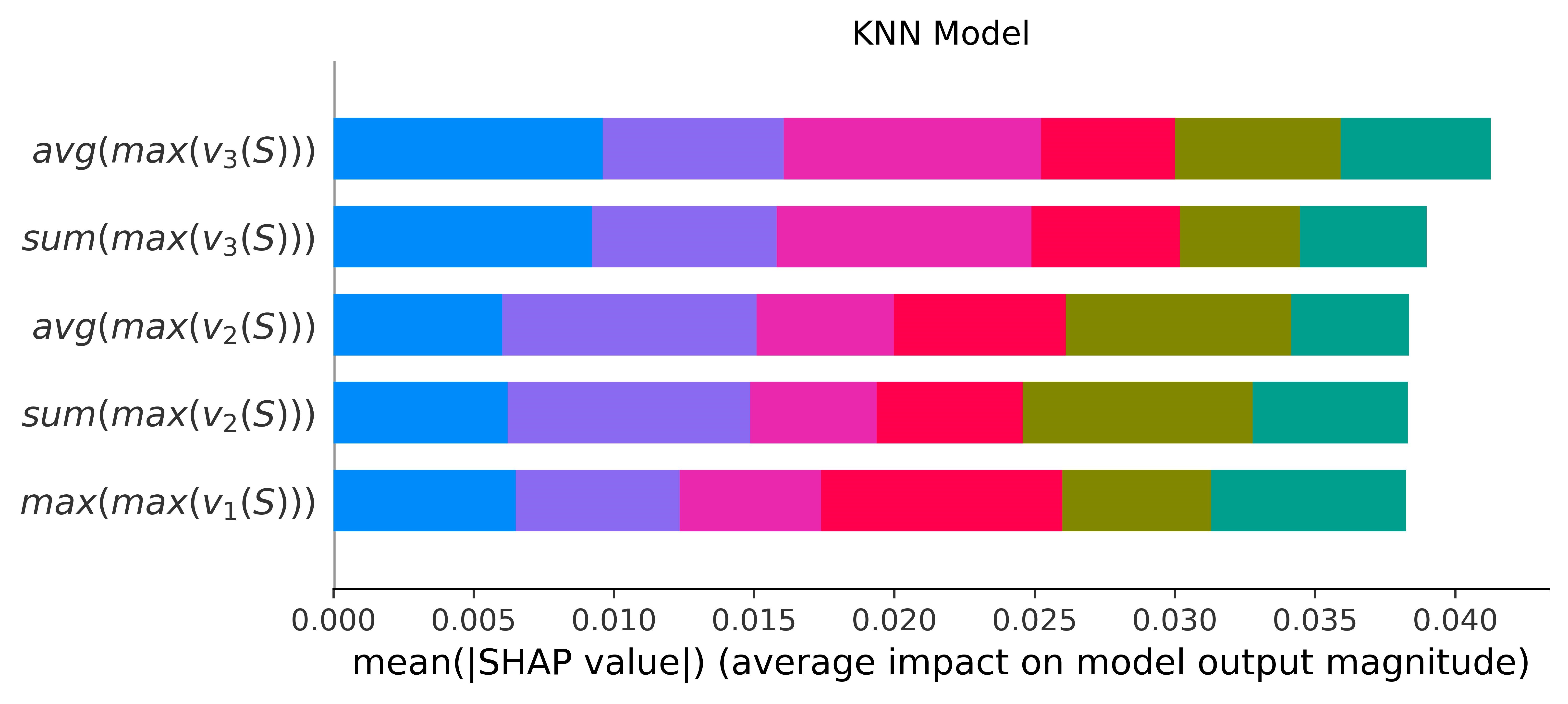}
         \label{KNN}
     \end{subfigure}
     \begin{subfigure}{.59\textwidth}
         \centering
         \includegraphics[width=\textwidth]{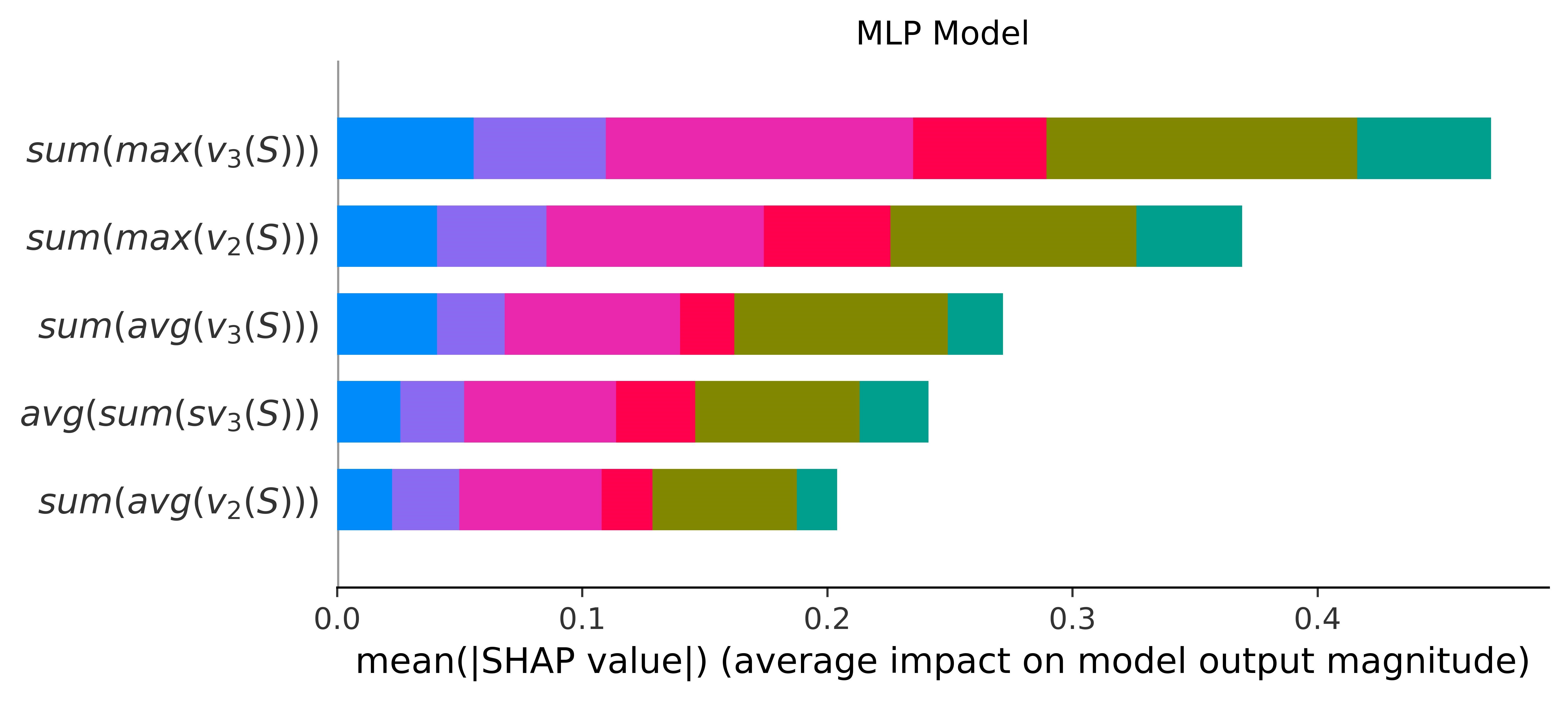}
         \label{MLP}
     \end{subfigure}
     \begin{subfigure}[b]{.59\textwidth}
         \centering
         \includegraphics[width=\textwidth]{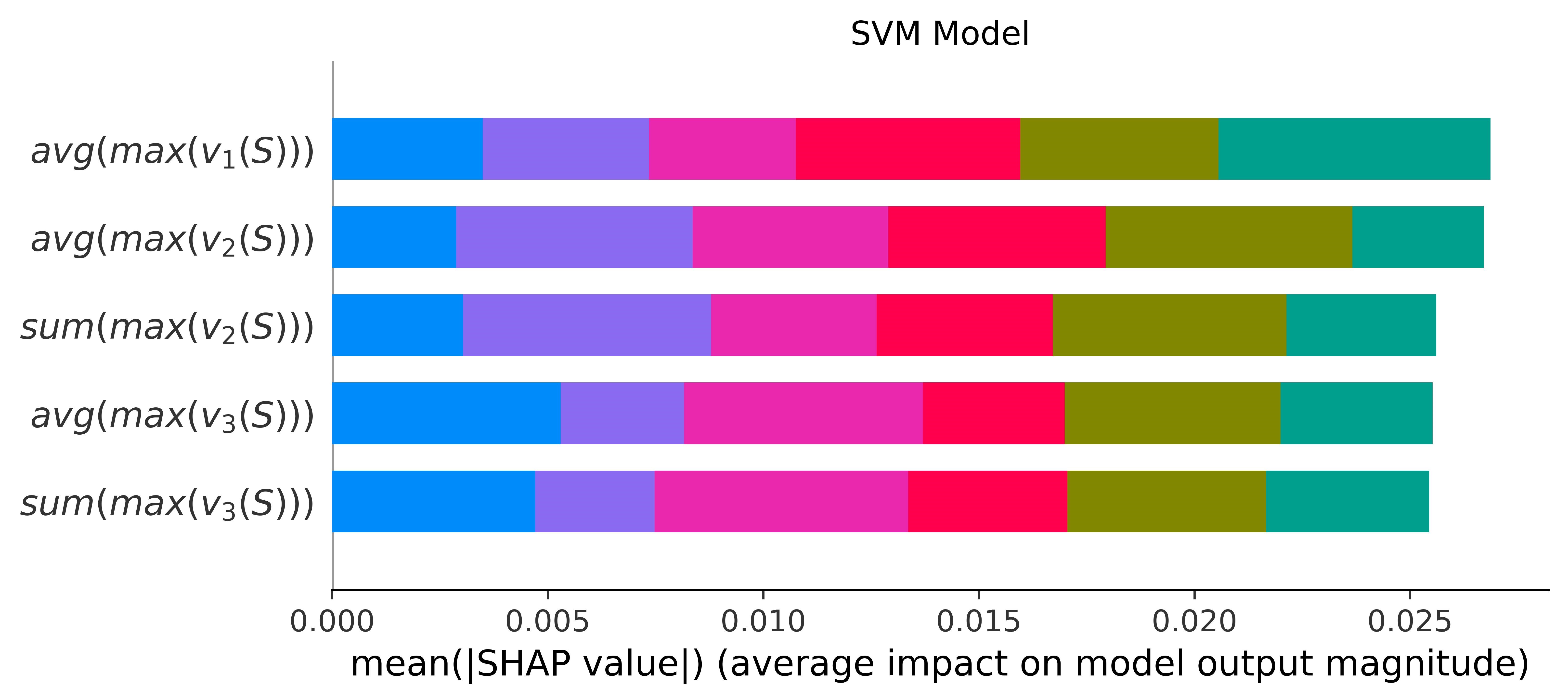}
         \label{SVM}
     \end{subfigure}
     \begin{subfigure}[b]{.59\textwidth}
         \centering
         \includegraphics[width=\textwidth]{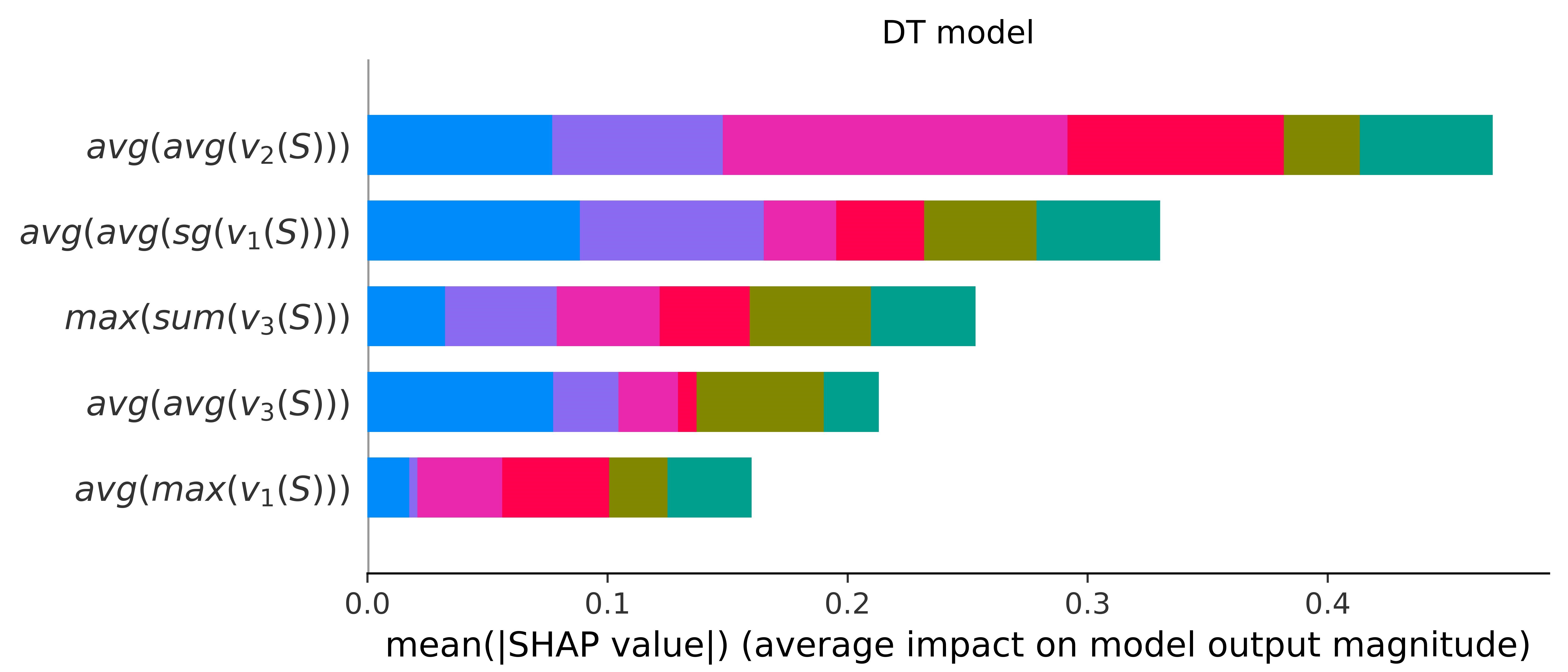}
         \label{DT}
     \end{subfigure}\\
     \begin{subfigure}[b]{.37\textwidth}
         \centering
         \includegraphics[width=\textwidth]{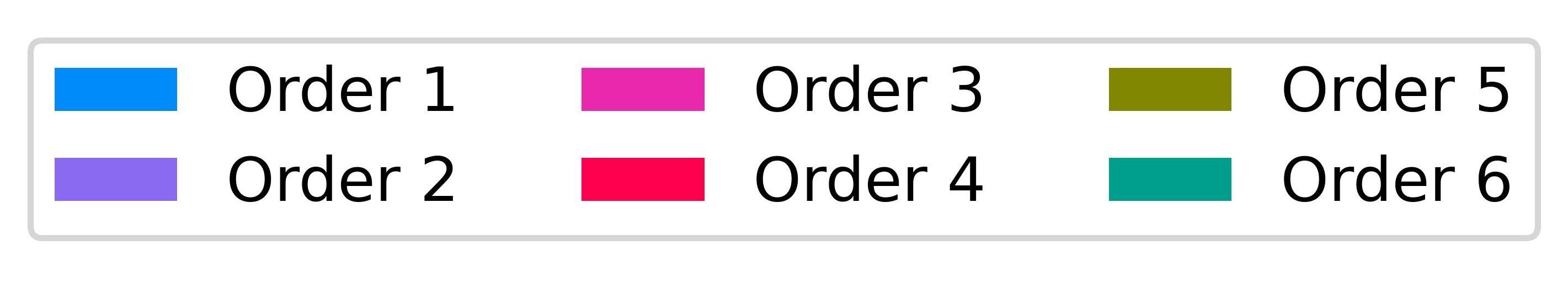}
         \label{ordering_label}
     \end{subfigure}
        \caption{The top five features for each model from aggregating over the 100 test points. Ordered by average impact on model output magnitude, and colored by output class.}
        \label{summary_top5_fig}
\end{figure}

Plots of the relative importance of features to each model by class are given in Figure \ref{summary_top5_fig}. These plots give a good insight into not only that feature's impact but also how a model trains and learns from the data. The KNN model, for example, creates little distinction between the impacts of the features. This is what we would expect from a KNN, because the model is looking for those instances which are most similar to the instance being evaluated to make a decision. It is the aggregate of the number of instances similar to a certain test point that will decide the output. This applies similarly to the SVM. For both models, this continues past the top five features. Specifically, the difference from the feature with the highest average impact on the model output magnitude to the twentieth feature is about 0.012 for the KNN, and 0.005 for the SVM. In contrast, the MLP and DT models show more distinction in the importance of these features:  decision trees in particular are known to place importance on a few features. To compare, the difference from the top to the fifth feature is about 0.28 for the ML.

Upon close examination of Figure \ref{summary_top5_fig}, some interesting phenomena may be observed. We see that often the impact on the output class for features of a certain variable is largest if that variable is last in the ordering of that output class. For example, the feature with the greatest impact on the KNN model is an $x_3$ feature, and Orderings 1 and 3 have the highest proportional impact which (referring back to Table \ref{tab: orderings}) are the orderings with $x_3$ last.  We see this for the $x_2$ features and the SVM model as well.  But the MLP model shows different behavior. Here, only $x_2$ and $x_3$ features are seen, and both Orderings 3 and 5 have the greatest impact.

\begin{table}[t]
    \centering
    \setlength{\tabcolsep}{20pt}
    \setlength\extrarowheight{2pt}
    \begin{tabular}{|c|c|c|c|c|}
    \hline
         & \textbf{KNN} & \textbf{MLP} \\ \hline
\textbf{1} & $avg(max(v_3(S)))$ & $sum(max(v_3(S)))$ \\ \hline
\textbf{2} & $sum(max(v_3(S)))$ & $sum(max(v_2(S)))$   \\ \hline
\textbf{3}  & $avg(max(v_2(S)))$ & $sum(avg(v_3(S)))$  \\ \hline
\textbf{4} & $sum(max(v_2(S)))$ & $avg(sum(sv_3(S)))$ \\ \hline
\textbf{5} & $max(max(v_1(S)))$ & $sum(avg(v_2(S)))$ \\ \hline
    \end{tabular}

    \qquad
    \newline

    \begin{tabular}{|c|c|c|c|c|}
    \hline
          & \textbf{SVM} & \textbf{DT} \\ \hline
\textbf{1} & $avg(max(v_1(S)))$  & $avg(avg(v_2(S)))$ \\ \hline
\textbf{2} & $avg(max(v_2(S)))$  & $avg(avg(sg(v_1(S)))$ \\ \hline
\textbf{3}  & $sum(max(v_2(S)))$ & $max(sum(v_3(S)))$ \\ \hline
\textbf{4} & $avg(max(v_3(S)))$ & $avg(avg(v_3(S)))$ \\ \hline
\textbf{5} & $sum(max(v_3(S)))$ & $avg(max(v_1(S)))$ \\ \hline
    \end{tabular}
    \caption{The top five features that each model found to have the highest average impact on model output magnitude.}
    \label{tab: top5_all_models}
\end{table}

\subsubsection{Features in human-designed heuristics}

The \texttt{Brown} heuristic \citep{Brown2004} uses three simple metrics to choose a variable ordering, all of which may be described in the language of Section \ref{subsubsec:features}:
\begin{itemize}
  \item $max(max(v_i(S)))$, 
  \item $max(max(sv_i(S)))$, and
  \item $sum(sum(sg(v_i(S))))$.
\end{itemize}

We cannot compare directly to the features of the other human-designed heuristics in Section \ref{subsec:prior_heuristics} as they all make use of expensive algebraic operations rather than using features from the input polynomials alone.

\section{Post SHAP analysis to identify the most relevant features}
\label{sec:explainability_analysis} 

SHAP produced for us a global analysis explaining the features that are important to each of the four ML models we experiment with.  We now work to create a unified ranking of the features across the four models for the purpose of creating new human-level heuristics in Section \ref{sec:new_heuristics}.

\subsection{Merging features that generate the same heuristic}

We aim for heuristics independent of particular variable names which are present in the ML features.  I.e. the prominence of $sum(sum(v_1(S)))$ in ML will lead us to the same heuristic as $sum(sum(v_3(S)))$. So we now aggregate the contribution of the features across the three variables, resulting in rankings like the one in Table \ref{tab: variables_merged} for each model.  This reduces the original 81 features to 27. We use $v_i(S)$ as a short notation to denote the combination across the three variables, i.e. 
$sum( sum( v_i(S))) := \textstyle \sum_{i=1}^3 sum( sum( v_i(S)))$.

\begin{table}[p]
    \centering
    \setlength{\tabcolsep}{20pt}
    \setlength\extrarowheight{2pt}
        \begin{tabular}{|c|c|}\hline
            \bfseries Feature Name & \bfseries \makecell{Summed \\ SHAP Value}
            \begin{filecontents*}{Datasets/Features_for_heuristics_long_MLP.csv}
FeatureNames,SummedSHAPValue
sum(max(v_i(S))),90.716
sum(avg(v_i(S))),57.566
sum(max(sv_i(S))),41.69
sum(sum(v_i(S))),36.67
avg(avg(sg(v_i(S)))),35.676
avg(sum(sv_i(S))),34.758
avg(max(sv_i(S))),30.981
avg(avg(v_i(S))),29.248
sum(sum(sv_i(S))),28.757
avg(avg(sv_i(S))),26.053
sum(sum(sg(v_i(S)))),25.164
sum(sg(avg(v_i(S)))),23.751
sum(avg(sv_i(S))),21.86
avg(sum(sg(v_i(S)))),21.495
sum(avg(sg(v_i(S)))),18.058
avg(sg(sum(v_i(S)))),16.486
max(max(v_i(S))),15.033
max(avg(v_i(S))),13.687
max(sum(sv_i(S))),13.443
max(max(sv_i(S))),12.029
avg(max(v_i(S))),11.864
max(max(sg(v_i(S)))),11.586
max(avg(sv_i(S))),11.559
max(sum(v_i(S))),11.389
max(sum(sg(v_i(S)))),10.451
avg(sum(v_i(S))),9.318
max(avg(sg(v_i(S)))),7.493
\end{filecontents*}
\csvreader[
            head to column names
            ]{Datasets/Features_for_heuristics_long_MLP.csv}{}{
            \\\hline $\FeatureNames$ & \SummedSHAPValue
            }
            \\\hline
    \end{tabular}
    \caption{Features important for the Multi Layer Perceptron after merging across the variables.}
    \label{tab: variables_merged}
\end{table}

Moreover, some features, even though they produce different values for the instances, share some relations. In particular, some of these features are proportional to each other (e.g. $sum(max(v_1(S)))$ is going to be a multiple of $(avg(max(v_1(S)))$) for every set of polynomials and therefore they can be seen theoretically to induce the same heuristics. This additional merging reduces the number of features further to 18. The ranking of such features for the MLP model is given in Table \ref{tab: fully_merged}.  Similar rankings are available for the other three models.

\begin{table}[!ht]
    \centering
    \setlength{\tabcolsep}{20pt}
    \setlength\extrarowheight{2pt}
        \begin{tabular}{|c|c|}\hline
            \bfseries Feature Name & \bfseries \makecell{Summed \\ SHAP Value}
            \begin{filecontents*}{Datasets/Features_for_heuristics_sum_MLP.csv}
FeatureNames,SummedSHAPValue
sum(max(v_i(S))),102.58
avg(avg(v_i(S))),86.814
sum(max(sv_i(S))),72.671
sum(sum(sv_i(S))),63.515
avg(avg(sg(v_i(S)))),53.735
avg(avg(sv_i(S))),47.913
sum(sum(sg(v_i(S)))),46.66
sum(sum(v_i(S))),45.988
avg(sg(sum(v_i(S)))),40.236
max(max(v_i(S))),15.033
max(avg(v_i(S))),13.687
max(sum(sv_i(S))),13.443
max(max(sv_i(S))),12.029
max(max(sg(v_i(S)))),11.586
max(avg(sv_i(S))),11.559
max(sum(v_i(S))),11.389
max(sum(sg(v_i(S)))),10.451
max(avg(sg(v_i(S)))),7.493
\end{filecontents*}
\csvreader[
            head to column names
            ]{Datasets/Features_for_heuristics_sum_MLP.csv}{}{
            \\\hline $\FeatureNames$ & \SummedSHAPValue
            }
            \\\hline
    \end{tabular}
    \caption{Features in Multi Layer Perceptron after merging those that would generate the same heuristic.}
    \label{tab: fully_merged}
\end{table}

\subsection{Creating a unified ranking}
\label{ssec:unifiedranking}

We now wish to create a combined feature ranking across the four models.  We will create this by viewing it as a voting problem, where each of the models has a different ranking for the features and we have to get a combined ranking.  A usual method to combine such rankings is the Borda Count: where for each ranking we assign a penalisation of 1 to the first, 2 to the second, and so on; at the end ranking the options from lowest penalization to highest. However, this method strongly penalises being very badly ranked for just one of the models. Meaning that a feature that is the most relevant for three of the models while being the most irrelevant for one of the models would not end up in a good position in the final ranking.  

Thus we use a modification, the Dowdall System, that assigns the first feature a reward of 1, the second a reward of 1/2, the third receives 1/3, and so on until, at the end ranking the options from higher reward to lowest.  See for example the work of \citet{Fraenkel2014} on these different voting systems.  We chose the Dowdall system because it allows merging the preferences without a big penalisation for being a very distant preference for one of the models, contrary to the Borda Count.

Table \ref{tab: top_features_overall} shows the features and their scores in our final ranking and their rewards from the Dowdall System. 
The most voted feature is the sum of the degrees of the polynomials in the set, which is precisely the feature used in \texttt{gmods} \citep{DelRio2022}, found through the study of the complexity analysis of CAD. In this ranking: the eleventh, eighteenth (bottom) and eighth most voted features are those used by the Brown heuristic \citep{Brown2004}.

The second most voted feature, the average across the polynomials of the average degree of the variable across the monomials, is a feature that has been never considered relevant in any prior work.

\begin{table}[!ht]
    \centering
    \setlength{\tabcolsep}{20pt}
    \setlength\extrarowheight{2pt}
        \begin{tabular}{|l|l|}\hline
            \bfseries Feature Name & \bfseries \makecell{Voted  \\ Score}
            \begin{filecontents*}{Datasets/score_of_features.csv}
Number,FeatureName,VotedScore
1,sum(max(v_i(S))),3.333
2,avg(avg(v_i(S))),2.167
3,sum(sum(v_i(S))),1.158
4,avg(avg(sg(v_i(S)))),1.15
5,sum(sg(sum(v_i(S)))),0.794
6,sum(max(sv_i(S))),0.787
7,avg(avg(sv_i(S))),0.583
8,sum(sum(sg(v_i(S)))),0.554
9,max(sum(v_i(S))),0.475
10,sum(sum(sv_i(S))),0.472
11,max(max(v_i(S))),0.467
12,max(sum(sg(v_i(S)))),0.3
13,max(avg(v_i(S))),0.245
14,max(max(sg(v_i(S)))),0.218
15,max(avg(sg(v_i(S)))),0.21
16,max(sum(sv_i(S))),0.209
17,max(avg(sv_i(S))),0.192
18,max(max(sv_i(S))),0.191
\end{filecontents*}
\csvreader[
            head to column names
            ]{Datasets/score_of_features.csv}{}{
            \\\hline $\FeatureName$ & \VotedScore
            }
            \\\hline
    \end{tabular}
    \caption{Voted score of merged and aggregated features across all models.   }
    \label{tab: top_features_overall}
\end{table}

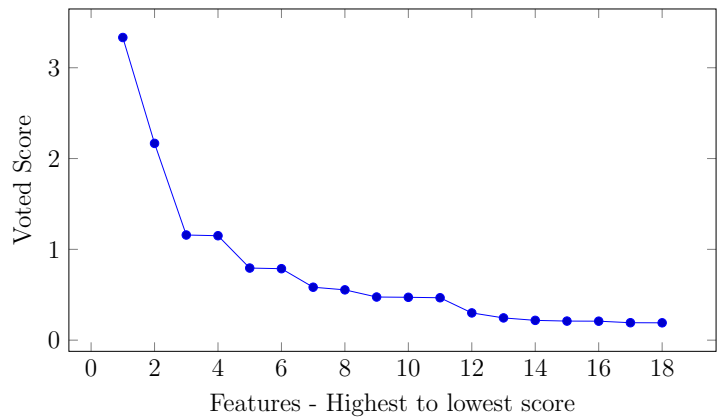
\begin{figure}
\centering
\begin{tikzpicture}[scale=0.8]
\begin{axis}[xlabel={Features - Highest to lowest score}, ylabel={Voted Score},x=15pt]
\addplot table [x=Number, y=VotedScore, col sep=comma] {Datasets/score_of_features.csv};
\end{axis}
\end{tikzpicture}
\caption{Plot of feature scores from Table \ref{tab: top_features_overall}}
\label{fig: feature scores}
\end{figure}

The top six features are chosen for further experimentation in the next section to find new human-level heuristics. We draw the line here as there is a big drop off and the lower features score more similarly to each other, i.e. the feature scores stabilize after the sixth score. This stabilization of the scores is illustrated in Figure \ref{fig: feature scores}.

\newpage

\section{New Human-Level Heuristics Motivated by XAI}
\label{sec:new_heuristics}

In this section, we will first introduce the methodology we use to evaluate a heuristic, and then experiment with new human-level heuristics motivated by Section \ref{sec:explainability_analysis}.  We will compare these against the current state-of-the-art heuristics for choosing the variable ordering for CAD. 

To evaluate the performance of these heuristics two things are needed: a set of meaningful benchmark examples and metrics to quantify the performance of the heuristics on them. We will use the benchmarks and metrics described recently by \citet{DelRio2022} for this purpose. 

\subsection{Benchmarks}

We source benchmarks from the three-variable problems in the QF\_NRA category of the SMT-LIB \citep{SMTLIB}.  These examples are all satisfiability problems and so do not represent the full application range of CAD which can also address quantifier elimination.  However, they do mostly emit from real applications making performance upon them particularly meaningful.  Common sources are problems emitting from the theorem prover MetiTarski \citep{Paulson2012}, attempts to prove termination of term-rewrite systems, verification conditions from Keymaera \citep{PQR09}, and parametrized generalizations of geometric problems, as well as problems emitting from economics \citep{MDE18} and biology \citep{BDEEGGHKRSW20}.   

A sign-invariant CAD is built using Maple \citep{CMXY09} for each example in each variable ordering: 5599 problems finished for at least one ordering before the time limit (60 seconds).  Of these, the authors select 1019 ``\emph{unique}'' problems where, as described in \citep[Section 4.1]{DelRio2022}, unique means the CAD tree structures built are not identical to another example for every ordering.  This benchmark merging is needed since some application sources in the SMT-LIB produce examples that differ from each other only slightly, leading to identical CAD computations, and thus potentially unfair evaluation.  The benchmarks used here are identical to those in \citep{DelRio2022}, which are stored in an open source dataset released for that paper\footnote{\url{https://doi.org/10.5281/zenodo.6750528}}.

\begin{figure}
    \centering
    \includegraphics[width=0.9\textwidth]{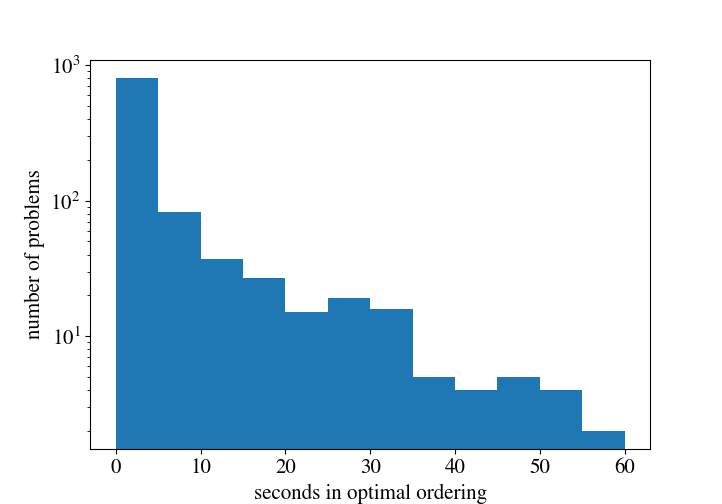}
    \caption{Histogram showing how the 1019 problems selected from the SMT-LIB for our experiments range in difficulty.  The horizontal axis has 5 second time segments and the vertical axis shows the number of problems whose optimal ordering took that level of computation time (in a logarithmic scale).}
    \label{fig:histogramD}
\end{figure}

The problems selected have a range of difficulties, as shown by the histogram in Figure \ref{fig:histogramD} which plots the number of problems against the CAD computation time of their optimal ordering.  The vertical axis (number of problems) has a logarithmic scale, showing that the dataset has much more easy problems.  There is a need to enlarge the SMT-LIB with more difficult problems.

\subsection{Evaluation metrics}

For evaluation metrics, we again follow \citet{DelRio2022} and use the three selected there as follows.
\begin{itemize}
    \item \texttt{Accuracy:} Percentage of benchmarks in which the optimal ordering (quickest in runtime) coincides with that chosen by the heuristic. 
\end{itemize}
This is the most commonly used tool in ML classification and was used to train the ML models earlier\footnote{See \citep{florescu2020improved} for an attempt to train ML partially with a different metric.}.  However, it is not ideally suited to our problem:  it considers as equally wrong a terrible ordering and an ordering that is slightly sub-optimal, even if the difference compared to the optimal ordering is so small that it could be due to computational noise.
\begin{itemize}
    \item \texttt{Total time:} Amount of time needed to create a CAD for each benchmark when choosing the variable ordering given by the heuristic. 
\end{itemize}
This metric corresponds most closely to the absolute performance achieved and would be most commonly used to rank a computer algebra implementation.  However, it will not take into account the chosen orderings for the smaller problems in our dataset and could allow a small number of large problems to distort findings.   Hence \citet{DelRio2022} proposed the next metric to address the shortcomings of the above: a metric that does not ignore the simple problems, but at the same time distinguishes between different non-optimal orderings.
\begin{itemize}
    \item \texttt{Markup:} The markup of the time needed by the heuristic ($t_{heuristic}$) with respect to the time using the optimal choice ($t_{optimal}$).  Defined as the average across all benchmarks of 
    \[
    \text{markup}=\frac{t_{heuristic}-t_{optimal}}{t_{optimal}+1}.
    \]
    The $+1$ is included in the denominator to reduce the impact that the computational noise can have on problems that only require tenths of seconds.
\end{itemize}
When a CAD could not be created within the time limit using the ordering chosen by a heuristic, the time of the heuristic was set in the dataset to twice the time limit.  These runtimes are also in the Zendodo data release for \citep{DelRio2022}.

\subsection{Existing heuristics}\label{subsec:existing_heuristics}
 
The prior state-of-the-art heuristic for choosing CAD variable ordering is \texttt{gmods}, as presented by \citet{DelRio2022}.  This is based on a metric  motivated by the complexity analysis of CAD. The analysis found that given a set of polynomials $S_n$ and the ordering $x_{i_n}\succ\dots\succ x_{i_1}$, the maximum number of cells that can be generated is proportional to 
\begin{equation}\label{eq: worst_case}
    \prod\limits_{j=1}^n sum(max(v_{i_j}(S_j))),
\end{equation}
where $S_{j-1}$ is the projection of $S_j$ with respect to $x_{i_j}$.

The heuristic \texttt{mods} was first proposed by \citet{DelRio2022} as a heuristic that computes the sets $S_1,\dots, S_{n-1}$ for each variable ordering and then chooses the ordering that minimises the worst case complexity, i.e. that minimises \eqref{eq: worst_case}.  
This performed very well, in the sense that it chooses good orderings. However, it is an expensive heuristic: the cost of computing the CAD projection phase for all the orderings was not compensated by the improved quality of the ordering it chooses, at least not for the dataset in question. This is illustrated later by the experimental results in Table \ref{tab:allvsall_existing_heuristics}, where \texttt{mods} took into account the additional cost of the heuristic while \texttt{free-mods} does not, we report on the performance of the hypothetical heuristic \texttt{free-gmods} for comparison purposes later.

To solve this issue \citet{DelRio2022} decided to avoid doing any projections for choosing the variable ordering except those utilised by the actual CAD. They proposed to pick the first variable of the ordering $x_{i_n}$ to be the one with minimal $sum(max(v_{i}(S_n)))$, choosing the one with the lowest index in case of a tie. Only then do they take a CAD projection of $S_n$ with respect to $x_{i_n}$ to obtain $S_{n-1}$, which is then studied to pick the second variable $x_{i_{n-1}}$ as the one with minimal $sum(max(v_{i}(S_{n-1})))$. This allows the heuristic to gather as much information as possible before choosing the next variable, continuing this process to determine the whole variable ordering.

This heuristic does not require any more CAD projections than those needed to run the CAD algorithm.  In this paper we refer to heuristics that follow this structure as ``\emph{greedy}'' (in the traditional algorithmic sense that they make a choice based on the local information).  All the heuristics introduced below are greedy\footnote{We note that \texttt{greedy-sotd} presented in \citep{Dolzmann2004} is not greedy in the sense of this paper because it makes a decision based on all the possible single projections for the next step (still less than the original \texttt{sotd} which computed all full projections).}.

Since the source benchmarks contain a bias towards certain orderings, the authors choose a variable randomly whenever there is a tie, instead of choosing the one with the lowest index or the first lexicographically which could allow that bias to effect the results.  

For many years, a well-known and widely used heuristic is that of \citet{Brown2004}. Denoted \texttt{Brown}, this uses three criteria, in turn, breaking ties with the subsequent ones.  It projects first the variable that minimizes:
\begin{itemize}
    \item $max(max(v_i(S)))$, (i.e. variable with lowest overall degree); \\ breaking ties with
    \item $max(max(sv_i(S)))$, (i.e. variable with the lowest total degree in the monomials containing it); breaking ties with
    \item $sum(sum(sg(v_i(S))))$, (i.e. number of terms containing the variable).
\end{itemize}
It is not specified what to do if there is a third tie: our implementation simply picks between the tied variables randomly.  Moreover, it is not specified whether the whole variable ordering is chosen at once or one variable at a time. In this paper, we implement the greedy variant:  i.e. we choose the first variable to be projected, and then use the projected polynomials to choose the next variable in the ordering.

Table \ref{tab:allvsall_existing_heuristics} evaluates these existing heuristics on our benchmark set.  We include also the performance of a virtual best heuristic (the hypothetical heuristic which always picks the optimal ordering) to show the best possible figures for the dataset, and a heuristic that picks at random.  As in \citep{DelRio2022} we find \texttt{gmods} to be the best of all metrics except accuracy: where it is beaten by \texttt{mods} but at a cost not worth paying for this dataset. 

Note that for some heuristics the number of problems completed is not a natural number. This is because the heuristic is not deterministic (due to the random choices to break ties) and the experiments are repeated 5000 times to obtain an average to account for this.

\begin{table}[H]
        \begin{tabular}{|l|l|l|l|l|l|}\hline
            \bfseries Name & \bfseries Accuracy & \bfseries Total time & \bfseries Markup
             & \bfseries \# Completed
            \begin{filecontents*}{Datasets/study_heuristics_guess__without_repetition__max_penalisation_inf__min_time_0.csv}
Name,TotalTime,Markup,NcellsMarkup,Terminating,NoSamples,Accuracy,30sTimeouts,60sTimeouts,FoundOutOf1,FoundOutOf2,FoundOutOf3
summaxdeg,7192.2,0.212,0.569,982.6,1019,0.563,27.018,9.36,0.702,0.981,0.945
avegavegdeg,7138.7,0.224,0.563,983.5,1019,0.544,27.004,8.528,0.682,0.988,0.959
sumsumdeg,7524.8,0.261,0.65,975.3,1019,0.549,34.516,9.144,0.665,0.967,0.957
avegavegsigndeg,8682.6,0.559,0.813,956.3,1019,0.535,47.028,15.648,0.563,0.887,0.912
sumsignsumdeg,10836.7,1.223,1.189,922.5,1019,0.45,79.126,17.418,0.507,0.81,0.81
summaxsvdeg,8771.7,0.563,0.85,956.5,1019,0.509,49.986,12.564,0.6,0.89,0.905
brown,7842.6,0.278,0.726,968.9,1019,0.553,41.008,9.14,0.647,0.945,0.939
mods,8137,0.127,0.383,979,1019,0.639,29,11,0.723,0.978,0.975
random,20797.3,4.034,2275.012,262.5,1019,0.167,744.167,12.333,0.167,0.333,0.5
free-mods,6637,0.566,0.383,990,1019,0.639,29,11,0.723,0.978,0.975
virtual-best,4822.7,0,0,1019,1019,1,0,0,1,1,1
0>1>2,6896.3,0.193,0.533,985.7,1019,0.567,26.599,6.658,0.695,0.988,0.962
0>4>2,6896.7,0.188,0.526,984.8,1019,0.583,27.565,6.662,0.707,0.974,0.963
\end{filecontents*}
\csvreader[
            head to column names,
            filter=\equal{\Name}{brown} \or \equal{\Name}{mods} \or \equal{\Name}{summaxdeg} \or \equal{\Name}{random} \or \equal{\Name}{virtual-best} \or \equal{\Name}{free-mods}
            ]{Datasets/study_heuristics_guess__without_repetition__max_penalisation_inf__min_time_0.csv}{}{

            \ifcsvstrequal{\Name}{summaxdeg}
            {\\\hline \texttt{gmods} & \Accuracy & \textbf{\TotalTime} & \textbf{\Markup}
            & \textbf{\Terminating}}
            {
            \ifcsvstrequal{\Name}{mods}
            {\\\hline \texttt{\Name} & \textbf{\Accuracy} & \TotalTime & \Markup
            & \Terminating}
            {
            \ifcsvstrequal{\Name}{brown}
            {\\\hline \texttt{Brown} & \Accuracy & \TotalTime & \Markup
            & \Terminating}
            {
            \ifcsvstrequal{\Name}{free-mods}
            {\\\hline\hline \texttt{\Name} & \Accuracy & \TotalTime & \Markup
            & \Terminating }
            {
            \\\hline \texttt{\Name} & \Accuracy & \TotalTime & \Markup
            & \Terminating
            }
            }
            }
            }
            }
            \\\hline
    \end{tabular}
    \caption{Evaluation of the existing heuristics to choose the variable orderings for CAD. In bold, the best measure of the metric out of all the heuristics. Note that after the double line, heuristics are not \label{tab:allvsall_existing_heuristics} }
\end{table}

\subsection{Creating new greedy heuristics from XAI identified features}\label{subsec:create_heuristic}

We will now build new heuristics solely based on a single feature from the earlier ML models.  We aim to see how powerful these features are when taking decisions on their own.

The explainability analysis in Section \ref{sec:explainability_analysis} found $sum( max( v_i(S)))$ to be the most impactful feature. A greedy heuristic \texttt{SumMaxV} based on this feature will select the variable for which this feature is minimal for each projection (and in the case of a tie one of the tied variables will be chosen randomly).

Consider for example the set $S_3=\{x_1x_2x_3-1, x_1^2-x_2^2x_3\}$; then \texttt{SumMaxV} will work as follows. First, the feature will be computed for each variable: 
\begin{align*}
   sum( max( v_1(S_3))) &= sum([1,2])=3,\\
   sum( max( v_2(S_3))) &= sum([1,2])=3,\\
   sum( max( v_3(S_3))) &= sum([1,1])=2.  
\end{align*} 
So $x_3$ is projected first because it has the minimal value.  The Lazard projection \citep{McCallum2016} of $S_3$ with respect to $x_3$ gives $\{x_1^3x_2-x_2^2, x_1x_2, x_2^2, x_1^2\}$. Factorizing and removing repeated factors gives $S_2=\{x_1^3-x_2, x_1, x_2\}$ and again we compute this feature for each of the remaining variables:
\begin{align*}
    sum( max( v_1(S_2))) &= sum([3,1,0]) = 4 \text{ and,} \\ sum( max( v_2(S_2)))&=sum([1,0,1])=2.
\end{align*} 
So $x_2$ is projected next because it has the minimal value. Thus heuristic \texttt{SumMaxV} chooses the ordering $x_3\succ x_2\succ x_1$.

\subsection{Performance of new single feature heuristics}

We now evaluate the new heuristics based upon the features that ranked in the top six in Table \ref{tab: top_features_overall} (selected as discussed at the end of Section \ref{ssec:unifiedranking}).  \texttt{AvgAvgSgV} is based upon feature $avg(avg(sg(v_i(S))))$, \texttt{SumMaxSV} is based upon $sum(max(sv_i(S)))$ and so on.  The results are shown in Table \ref{tab:allvsallInf0}.  

It is not a coincidence that the statistics for \texttt{SumMaxV} in Table \ref{tab:allvsallInf0} are the same as for \texttt{gmods} in Table \ref{tab:allvsall_existing_heuristics}:  \texttt{gmods} can be viewed as single feature heuristic based on $sum( max( v_i(S)))$.

\begin{table}[ht]
        \begin{tabular}{|l|l|l|l|l|l|}\hline
            \bfseries Name & \bfseries Accuracy & \bfseries Total time & \bfseries Markup
             & \bfseries \# Completed
            \csvreader[
            head to column names,
            filter=\equal{\Name}{sumsumdeg} \or \equal{\Name}{avegavegdeg} \or \equal{\Name}{sumsignsumdeg} \or \equal{\Name}{avegavegsigndeg} \or \equal{\Name}{summaxdeg} \or \equal{\Name}{summaxsvdeg}
            ]{Datasets/study_heuristics_guess__without_repetition__max_penalisation_inf__min_time_0.csv}{}{

            \ifcsvstrequal{\Name}{sumsumdeg}
            {\\\hline \texttt{SumSumV} & \Accuracy & \TotalTime & \Markup
            & \Terminating}
            {
            \ifcsvstrequal{\Name}{summaxdeg}
            {\\\hline \texttt{SumMaxV} & \textbf{\Accuracy} & \TotalTime & \textbf{\Markup}
            & \Terminating}
            {
            \ifcsvstrequal{\Name}{avegavegdeg}
            {\\\hline \texttt{AvgAvgV} & \Accuracy & \textbf{\TotalTime} & \Markup
            & \textbf{\Terminating}}
            {
            \ifcsvstrequal{\Name}{sumsignsumdeg}
            {\\\hline \texttt{SumSgSumV} & \Accuracy & \TotalTime & \Markup
            & \Terminating}
            {
            \ifcsvstrequal{\Name}{summaxsvdeg}
            {\\\hline \texttt{SumMaxSV} & \Accuracy & \TotalTime & \Markup
            & \Terminating}
            {
            \ifcsvstrequal{\Name}{avegavegsigndeg}
            {\\\hline \texttt{AvgAvgSgV} & \Accuracy & \TotalTime & \Markup
            & \Terminating}
            {
            \\\hline \texttt{\Name} & \Accuracy & \TotalTime & \Markup
            & \Terminating}
            }
            }
            }
            }
            }
            }
            \\\hline
    \end{tabular}
    \caption{Evaluation metrics for the new heuristics to choose the variable orderings for CAD. In bold, the best measure of the metric out of all the heuristics.  \label{tab:allvsallInf0} }
\end{table}

The table shows there are two heuristics features that stand out: \texttt{SumMaxV} (\texttt{gmods}) and \texttt{AvgAvgV}. These achieve similar results, and to chose the best feature would depend on the metric used: neither dominates the other. 

\subsection{New heuristics of triples of features identified by SHAP}
\label{subsec:triplets}

We have now seen a number of single-feature heuristics doing well, without a clear winner. We note that even the best of these (\texttt{gmods}) still relies on a random tie-breaker for 25\% of the problem instances, indicating it does not alone have access to enough information.  So we will next combine these single features into triples, using subsequent ones as tie-breakers. This approach is the same as that used by the heuristic of \citet{Brown2004}. This should give a more informed choice than selecting a random variable and is still certainly at a human level in size and complexity.

We will denote these heuristics based on multiple features by concatenating their names with a $>$, where smaller indicates that it is used as a tie-breaker for the bigger features. Following this notation the \texttt{Brown} heuristic would be denoted as \texttt{MaxMaxV>MaxMaxSV>SumSumSgV}.

There are 120 possible ordered triples that can be created from the top six features of Table \ref{tab: top_features_overall}.  For each we create a heuristic that chooses the variable that minimises the first feature, breaking ties using the second feature and the third if necessary, and finally breaking ties by choosing a random variable if there is a tie for all three features.  

Performance statistics for the two triples that are best in one of the metrics are given in Table \ref{tab:allvsallInf02}, along with \texttt{Brown} that makes choices with a similar quantity of information. 
 
\begin{table}[ht]
        \begin{tabular}{|l|l|l|l|l|l|}\hline
            \bfseries Name & \bfseries Accuracy & \bfseries Total time & \bfseries Markup
             & \bfseries \# Completed
            \begin{filecontents*}{Datasets/study_heuristics_guess__without_repetition__max_penalisation_inf__min_time_0-capB.csv}
Name,TotalTime,Markup,NcellsMarkup,Terminating,NoSamples,Accuracy,30sTimeouts,60sTimeouts,FoundOutOf1,FoundOutOf2,FoundOutOf3
summaxdeg,7192.2,0.212,0.569,982.6,1019,0.563,27.018,9.36,0.702,0.981,0.945
avegavegdeg,7138.7,0.224,0.563,983.5,1019,0.544,27.004,8.528,0.682,0.988,0.959
sumsumdeg,7524.8,0.261,0.65,975.3,1019,0.549,34.516,9.144,0.665,0.967,0.957
avegavegsigndeg,8682.6,0.559,0.813,956.3,1019,0.535,47.028,15.648,0.563,0.887,0.912
sumsignsumdeg,10836.7,1.223,1.189,922.5,1019,0.45,79.126,17.418,0.507,0.81,0.81
summaxsvdeg,8771.7,0.563,0.85,956.5,1019,0.509,49.986,12.564,0.6,0.89,0.905
Brown,7842.6,0.278,0.726,968.9,1019,0.553,41.008,9.14,0.647,0.945,0.939
mods,8137,0.127,0.383,979,1019,0.639,29,11,0.723,0.978,0.975
random,20797.3,4.034,2275.012,262.5,1019,0.167,744.167,12.333,0.167,0.333,0.5
free-mods,6637,0.566,0.383,990,1019,0.639,29,11,0.723,0.978,0.975
virtual-best,4822.7,0,0,1019,1019,1,0,0,1,1,1
0>1>2,6896.3,0.193,0.533,985.7,1019,0.567,26.599,6.658,0.695,0.988,0.962
0>4>2,6896.7,0.188,0.526,984.8,1019,0.583,27.565,6.662,0.707,0.974,0.963
\end{filecontents*}
\csvreader[
            head to column names,
            filter=\equal{\Name}{Brown} \or \equal{\Name}{0>1>2} \or \equal{\Name}{0>4>2}
            ]{Datasets/study_heuristics_guess__without_repetition__max_penalisation_inf__min_time_0-capB.csv}{}{

            \ifcsvstrequal{\Name}{0>1>2}
            {\\\hline \texttt{T1} & \Accuracy & \textbf{\TotalTime} & \Markup
            & \textbf{\Terminating}}
            {

            \ifcsvstrequal{\Name}{0>4>2}
            {\\\hline \texttt{T2} & \textbf{\textbf{\Accuracy}} & \TotalTime & \textbf{\Markup}
            & \Terminating}
            {
            \\\hline \texttt{\Name} & \Accuracy & \TotalTime & \Markup
            & \Terminating
            }
            }
            }
            \\\hline
    \end{tabular}
    \caption{Evaluation metrics for the different heuristics to choose the variable orderings for CAD. In bold, the best measure of the metric out of all the heuristics. \texttt{T1}$=$\texttt{SumMaxV>AvgAvgV>SumSumV} and \texttt{T2}$=$\texttt{SumMaxV>SumSumSgV>SumSumV} \label{tab:allvsallInf02} }
\end{table}

\begin{figure}[ht]
    \centering
    \includegraphics[width=0.8\textwidth]{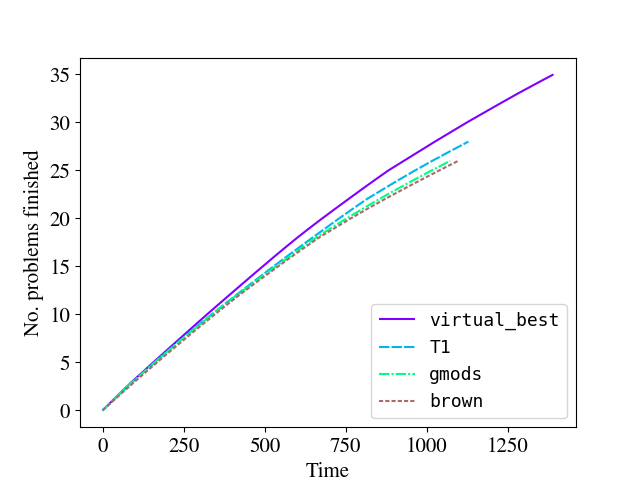}
    \caption{Survival plot for best-existing heuristics and new heuristics proposed in the problems that take more than 30 seconds for all the orderings.}
    \label{fig: survival_plot}
\end{figure}

Table \ref{tab:allvsallInf02} shows that the heuristic \texttt{T1}$=$\texttt{SumMaxV>AvgAvgV>SumSumV} wins in two of the four metrics: total time and the number of problems completed within the time limit. \texttt{T1} relies on the random tie-breaker for only 5\% of the problem instances, far less than any of the single feature heuristics, as expected. Note that this heuristic is created from the top three features XAI suggested features, in the order they appear in the voting system.
\begin{itemize}
    \item $sum(max(v_i(S)))$, (i.e. variable with the lowest degree in the product of the polynomials);  breaking ties with
    \item $avg(avg(v_i(S)))$, (i.e. variable with the lowest average of average degree in the polynomials); breaking ties with
    \item $sum(sum(v_i(S)))$, (i.e. variable with the lowest sum of all its degrees).
\end{itemize}
This triple also outperforms all the singleton heuristics, and its performance gets really close to the performance of the hypothetical heuristic \texttt{free-gmods}, shown in Table \ref{tab:allvsall_existing_heuristics}.

In Figure \ref{fig: survival_plot} we give a survival plot comparing the triple T1 to the prior state-of-the-art and the virtual best.  We do this just for the hardest problems in the dataset (those taking more than 30 seconds in all orderings) to allow for a clearer visualisation of the differences.  The figure shows the leap that existed between the prior state-of-the-art and the best possible in theory, and that the new XAI informed triple bridges a good deal of that gap. 

\section{Conclusions}
\label{sec:conclusions}

\subsection{Summary}

In this paper, we have devised a new state-of-the-art human-level heuristic for choosing a CAD variable ordering, at least in three variable problems.  Previous approaches to the problem were of two types: either an expert handpicking features to create a human-designed heuristic; or an ML model computing with many features under the hood. We presented an intermediate approach where we start with the ML model but then use XAI to select important features, which may then be used later independently of any ML technology.   

We note that SHAP was able to find the feature identified recently by \citet{DelRio2022} that was used to form the prior state-of-the-art heuristic; and identified also the tools of the heuristic in \citet{Brown2004}. But it also found an important feature not previously considered by experts.

\subsection{Future work}

There is of course scope for future work.  Using multiple features clearly improved the results, and there are certainly others ways we could consider combining them. Perhaps this combination itself could also be ML informed? It would also be interesting to explore alternative explainable AI methods, especially ML models that are inherently explainable, rather than the post-hoc explainability analysis of SHAP.  We also acknowledge the need to experiment with other datasets including problems with more variables.  

The experiments presented in this paper are for selecting CAD variable ordering. However, the methodology could be applied almost directly to variable ordering choices for other algorithms, and could be adapted to many algorithm optimisation problems in symbolic computation and beyond. We have shown that ML has a role to play in symbolic computation, even for developers who prefer not to include a reliance upon it in their final code.  It can be used as a tool to guide and develop algorithms at a human level. We thus hope and expect to see other applications of XAI to inform decisions in CASs and optimise symbolic computation algorithms in the coming years.

\section*{Acknowledgements}

Matthew England acknowledges the support of UKRI EPSRC Grant EP/T015748/1, ``\emph{Pushing Back the Doubly-Exponential Wall of Cylindrical Algebraic Decomposition}'' (DEWCAD).  Tereso del Río and Lynn Pickering acknowledge the Coventry University Research Excellence grant that allowed them to work together in person on this paper. Lynn Pickering acknowledges the support of the Rindsberg Fellowship from the University of Cincinnati, the Ohio Space Grant Consortium Research Fellowship, and a University of Cincinnati International Study Abroad Scholarship that allowed her to spend a semester at Coventry University.

We than AmirHosein Sadeghimanesh for helpful feedback, and the anonymous reviewers whose comments greatly improved this paper.

\section*{Research Data Statement}

The code and data supporting the research in this paper are freely available from Zenodo: \url{https://doi.org/10.5281/zenodo.8229298}

\bibliographystyle{elsarticle-harv}
\bibliography{bib.bib}

\end{document}